\documentclass[english]{iopart}
			   
\begin{document}
				   
\title[Fuchs versus Painlev\'e] 
{\Large
 Fuchs versus Painlev\'e
}

\author{S. Boukraa$^\dag$,  S. Hassani$^\S$,
J.-M. Maillard$^\ddag$, B. M. McCoy$^\P$,
 J.-A. Weil$^\star$ and N. Zenine$^\S$}
\address{\dag Universit\'e de Blida, LPTHIRM and D\'epartement
 d'A{\'e}ronautique,
 Blida, Algeria}
\address{\S  Centre de Recherche Nucl\'eaire d'Alger, \\
2 Bd. Frantz Fanon, BP 399, 16000 Alger, Algeria}
\address{\ddag\ LPTMC, Universit\'e de Paris 6, Tour 24,
 4\`eme \'etage, case 121, \\
 4 Place Jussieu, 75252 Paris Cedex 05, France} 
\address{\P Institute for Theoretical Physics,
State University of New York,
Stony Brook, USA}
\address{$\star$  XLIM, Universit\'e de Limoges,
123 avenue Albert Thomas,
87060 Limoges Cedex, France} 
\ead{boukraa@mail.univ-blida.dz, maillard@lptmc.jussieu.fr, maillard@lptl.jussieu.fr, 
mccoy@max2.physics.sunysb.edu, jacques-arthur.weil@unilim.fr, njzenine@yahoo.com}

\begin{abstract}
 
We  briefly\footnote[2]{This paper is for the
Special issue on Symmetries and Integrability of Difference Equations (SIDE VII).}, 
recall 
the Fuchs-Painlev\'e elliptic representation
of Painlev\'e VI. We then show that the 
 polynomiality of the expressions 
 of the  correlation functions (and form factors) 
 in terms of the complete elliptic integral of the first and second kind, 
 $\, K$  and $\, E$, is a straight consequence
 of the fact that the differential operators corresponding to
the entries of Toeplitz-like determinants, are equivalent to the second order
 operator $\, L_E$ which has $\, E$ as solution (or, for 
off-diagonal correlations
to the direct sum of  $\, L_E$ and $\, d/dt$). We show that this 
 can be generalized, mutatis mutandis, to the
 anisotropic Ising model. The singled-out  second order 
linear differential operator $\, L_E$ is replaced
 by an isomonodromic system of two third-order linear
 partial differential operators
 associated with $\, \Pi_1$, the Jacobi's form of the 
complete elliptic integral of the third kind (or equivalently 
 two second order linear partial differential operators associated
 with Appell functions, where one of these operators
can be seen as a deformation of $\, L_E$).
We  finally explore the generalizations, to
 the anisotropic Ising models, of the links we made, 
in  two previous papers, between Painlev\'e 
non-linear  ODE's, Fuchsian linear ODE's and elliptic curves. 
In particular the elliptic representation of Painlev\'e VI
 has to be generalized 
to an ``Appellian'' representation of Garnier systems.

\end{abstract}

\vskip .1cm

\vskip .2cm
\noindent {\bf AMS Classification scheme numbers}: 
33E17, 33E05, 33Cxx,  33Dxx, 14Exx, 14Hxx, 34M55, 47E05,
 34Lxx, 34Mxx, 14Kxx
\vskip .1cm
 {\bf Keywords}:  sigma form of Painlev\'e VI, two-point 
correlation functions of the Ising model,  
Fuchsian linear differential equations,  
 complete elliptic integrals of the first, second and third kind, elliptic 
representation of Painlev\'e VI, Appell Hypergeometric
 functions, Garnier systems, Schlesinger systems.

\vskip .1cm

\vskip .1cm

\vskip .1cm

\section{Introduction}
\label{intro}

 In a previous paper~\cite{PainleveFuchs} we have shown that 
the {\em  diagonal} two-point correlation functions of the 
square Ising model are simultaneously solutions 
of a non-linear differential equation associated
 with (the sigma form of\footnote[3]{More precisely the
 $\, \sigma$ associated with the
log-derivative of the diagonal two-point correlation function is solution of
 the sigma form of Painlev\'e VI.}) Painlev\'e VI, and solutions 
of Fuchsian linear differential equations. In a following
 paper~\cite{Holo} we have also
shown that some one-parameter 
$\, \lambda$-extension of the diagonal two-point correlation
 functions (which also satisfy the same 
non-linear differential equations associated with the sigma form of
 Painlev\'e VI) are such that their coefficients in the 
$\, \lambda$-series, the so-called ``form factors'', also verify, 
in a rather unexpected
way, Fuchsian linear differential equations~\cite{Holo}. 
More precisely, introducing the  second order differential operator $\, L_E$, 
associated~\cite{PainleveFuchs,Holo} with the
 complete elliptic integral of the second kind $\, E$,
 the linear differential operators forming 
these two sets of  Fuchsian  ODE's (for the two-point
 correlation functions and for the form factors)
were seen to be equivalent\footnote[8]{In the sense of
the equivalence of linear differential 
operators~\cite{Singer,PutSinger}.} to direct sums of linear differential 
operators equivalent to symmetric powers
 of  $\, L_E$, or simply (for diagonal correlations) to symmetric powers of  $\, L_E$. 
As a consequence, the two-point correlation functions,
 as well as the previously mentioned form factors~\cite{Holo},
are polynomials expressions of $\, K$ and $\, E$, the 
complete elliptic integral of the first and second kind.
 These results underline the key role played by the 
second order differential operator $\, L_E$,
and this can be seen  to be in perfect agreement with the 
{\em elliptic representation} of the Painlev\'e VI equations~\cite{Holo}. A surprisingly
 large amount of informations on correlation functions
and form factors, is thus  ``encapsulated'' in this second order  
differential operator $\, L_E$. This suggests two sets of work to be performed.
 
First, we revisit our two previous papers~\cite{PainleveFuchs,Holo} in order to show
 that the results displayed in these two
 papers are, in the case of the {\em isotropic}\footnote{Or 
the anisotropic Ising model but {\em only 
for diagonal} two-point correlation functions: the  diagonal 
 correlations are only a function of
 {\em only one variable}, the modulus $\, k$, and not of the 
anisotropy of the model.} Ising model,
 direct consequences of the natural occurence of
 the second order linear differential operator $\, L_E$.

 Secondly, we try to generalize these calculations
 to the {\em anisotropic} Ising model, to see if a similar ``scheme'' can be 
generalized, mutatis mutandis. In the  {\em anisotropic} Ising model,
 the linear differential
 operators for the  two-point correlation functions,
and the form factors, could be, again, direct sums of linear differential 
operators equivalent to symmetric powers of 
new linear partial differential systems  to be
 discovered. As a byproduct, the finding of these new linear partial 
 differential operators, generalizing $\, L_E$, would indicate the
 proper generalization of the elliptic representation of Painlev\'e VI, for
the anisotropic Ising model (and more generally, for integrable
 lattice models with a canonical elliptic parametrization of
their Yang-Baxter equations, like the Baxter model). What is 
the ``natural'' generalization
 of the elliptic representation of Painlev\'e VI for off-diagonal
correlation functions, and their $\, \lambda$-extensions, for
 the {\em anisotropic} Ising model ?
 Should we introduce higher order Painlev\'e ODE's 
(in analogy to the higher order KdV generalization of KdV) ?  Should we
 consider Garnier systems~\cite{Garnier}, or even, more general
 Schlesinger systems~\cite{Schlesinger,Schlesinger2} ? 

This paper is organized as follows. We will, briefly,
 recall the Fuchs elliptic representation
of Painlev\'e VI, and then show that correlation functions 
being polynomial expressions in the complete elliptic integrals 
$\, E$ and $\, K$, is a straight consequence
 of the fact that the linear differential operators corresponding to
the entries of some Toeplitz-like determinants, are equivalent
 to the second order
 operator $\, L_E$ (or, for off-diagonal correlations to
 direct sums of  $\, L_E$ and $\, d/dt$).
 We will show that these previous calculations
 can be generalized, mutatis mutandis, to the
 {\em anisotropic} Ising model, the singled-out  second order 
linear differential operator $\, L_E$ being 
replaced by a system of two (isomonodromic) {\em third order} linear partial
 differential operators corresponding to the 
elliptic integral of the third kind $\, \Pi$, or, equivalently, 
two (isomonodromic) {\em second order} linear partial differential 
operators corresponding to {\em Appell functions}.
We will finally explore the generalizations, to
 the anisotropic Ising models, of the links we made 
in the two previous papers, between Painlev\'e
 non-linear  ODE's, Fuchsian linear ODE's~\cite{PainleveFuchs},
and elliptic curves~\cite{Holo}. We will suggest that
the elliptic representation of Painlev\'e VI has to be generalized 
to an ``Appellian'' representation of {\em Garnier systems}.

\section{About Painlev\'e VI}
\label{Painl}

\subsection{Ising model and the sigma form of Painlev\'e VI}
\label{sigmaform}

For concreteness we first recall the specific sigma form of
 Painlev\'e VI  obtained
by Jimbo and Miwa~\cite{jim-miw-80} for the diagonal two-point
 Ising correlation $\, C(N,N)$:
\begin{eqnarray}
\label{jimbo-miwa}
&&\Bigl( t\, (t-1) \sigma^{''} \Bigr)^2\, =\,\,   \\
&&\quad \quad N^2 \cdot \Bigl( (t-1) \sigma^{'} -\sigma \Bigr)^2\,\, 
 -4\,\sigma^{'}
\cdot  \Bigl((t-1)\sigma^{'}-\sigma -1/4   \Bigr)
\cdot  \Bigl(t\sigma^{'}-\sigma   \Bigr). \nonumber 
\end{eqnarray}

The diagonal correlation $C_N=\, C(N,N)$ is related to 
 $\sigma$, for $T >T_c$, by~\cite{PainleveFuchs}
\begin{eqnarray}
\label{sigma-bas}
&&\sigma(t)\,=\,\, \,t\cdot (t-1) \cdot {\frac{d}{dt}}\log(C_N)\,
-{{1} \over {4}}\nonumber\\
&&{\rm with} \quad  \quad \quad \quad t\, 
=\, \, \Bigl( \sinh(2J_v/kT)\cdot \sinh(2J_h/kT) \Bigr)^2 <1
\end{eqnarray}
and, for $\, T <T_c$, by
\begin{eqnarray}
\label{sigma-haut}
&&\sigma(t)\,=\,\,\, t\cdot  (t-1)  \cdot{\frac{d}{dt}}\log(C_N)\,\,
-{{t} \over {4}}\nonumber\\
&&{\rm with} \quad  \quad \quad \quad t\, =\, \, 
\Bigl( \sinh(2J_v/kT)\cdot \sinh(2J_h/kT) \Bigr)^{-2} <1
\end{eqnarray}
where the  variable  
$J_v$ ($J_h$) is the Ising model vertical (horizontal) coupling constant.

\subsection{Fuchs-Painlev\'e elliptic representation of Painlev\'e VI}
\label{ellirepr}

Let us introduce $\, K$ and $\, E$, the complete elliptic integral
of the first kind and of the second kind
that we multiply\footnote[3]{In maple's notations, for
 $ t \, = \, k^2$ ($k$ is the modulus), 
$\, K(t) \, = \, K(k^2)$ in (\ref{E}) 
 reads : $\, hypergeom( [1/2,1/2],[1],t)$  
 $\,= \,  2/\pi \cdot EllipticK(k)$, but reads $ 2/\pi \cdot EllipticK[k^2]$ 
in Mathematica.} by $2/\pi$ in order
 to have series with integer coefficients: 
\begin{eqnarray}
\label{E}
 K(t) \, = \, \,\, {_2}F_1 \left( 1/2, 1/2; 1; t \right), \quad  
\quad  E(t)\,=\,\, {_2}F_1 \left( 1/2, -1/2; 1; t\right). 
\end{eqnarray}

Let us also introduce the second order differential operator corresponding to
 $\, E$ ($Dt\, $ denotes the derivative with respect
 to $\, t$: $Dt\, = \, \, d/dt$) :
\begin{eqnarray}
\label{LE}
L_E \, = \, \, \, \, \, Dt^2\, 
\, +{\frac { Dt}{t}}\, \, - \,{\frac {1}{ 4 \, (t-1)\, t }}.
\end{eqnarray}

In order to understand the key role played by $\, L_E$,
let us first
 recall (see~\cite{man}, or for a review~\cite{Guzzetti}) the
 so-called Fuchs-Painlev\'e
``{\em elliptic representation}'' of
 Painlev\'e VI. This  elliptic representation
of Painlev\'e VI amounts to seeing Painlev\'e VI as a 
``deformation'' (see equation (33) in~\cite{Guzzetti})  of the 
hypergeometric linear differential equation 
 associated with the second order linear differential operator :
\begin{eqnarray}
\label{Lfuchs}
{\cal L} \, = \, \, \, \, (1-t) \,  t \cdot Dt^2 \,  \,
 + (1-2\, t) \cdot Dt \, \,  -{{1} \over {4}}. 
\end{eqnarray}
One easily verifies that this linear differential operator 
 has the complete elliptic integral of the first kind $\, K$
as solution. We will  denote  $ L_K$ the second order operator defined by
 $\,{\cal L} \, = \, \,  (1-t) \,  t \cdot L_K$.
 The operator  $\,{\cal L} \, $ is actually 
equivalent (in the sense of the equivalence
of linear differential operators~\cite{Singer,PutSinger}) with $\, L_E$ :
\begin{eqnarray}
\label{Lfuchsequiv}
L_E \cdot \Bigl(2\, (t-1)\, t \cdot Dt \, + t \, -1 \Bigr) \, \, = \, \, \, \, 
\Bigl( -2\cdot Dt \, -{{3} \over {t}}  \Bigr) \cdot {\cal L}.
\end{eqnarray}

This deep relation between {\em elliptic curves and Painlev\'e VI}
explains the occurrence of Painlev\'e VI on the isotropic Ising model,
and on other lattice Yang-Baxter integrable models
which are canonically parametrized in term of {\em elliptic functions} 
(like the eight-vertex Baxter model, the RSOS models, 
see for instance~\cite{Manga}). 

\section{ Fuchsian linear ODE's for Ising two-point correlations}
\label{first}

In this section, we prove 
the polynomiality of the two-point correlation functions 
in $\, E $ and $\, K $, in a way that underlies 
  differential algebra and  the equivalence of  linear differential operators
(see (\ref{equiv}) below), since this approach
 can easily (but tediously) be generalized to
the anisotropic Ising model (see section (\ref{generalis}) below). 

For the isotropic square Ising model we consider
 the regime $\, T \, > \, T_c$, and
 we use the same notations\footnote[5]{We apologize for
possible repetition of material appearing in this section
 and some relevant parts of~\cite{PainleveFuchs}. 
We consider that the reader may not be familiar with differential algebra
concepts, in particular
 the notion of equivalence of linear differential operators.} as in~\cite{PainleveFuchs,Holo},
namely $\, s \, = \, \sinh(2\, K)$ and
 $\, t \, = \, \, k^2 \, = \, \, s^4$ ($k$ is the modulus of the
 elliptic functions). We will use, alternatively, the
two variables $\, t \,$ and $\, s \,$ (according to
 the quantity we study: for off-diagonal two-point correlations
the  $\, s$ variable is better suited).
The diagonal two-point correlation functions of the square Ising model
$\, C(N,N)$, and its dual $\, C^*(N,N)$, can be
calculated from Toeplitz determinants~\cite{kau-ons-49,mo-po-wa-63,mcc-wu-73}:
\begin{eqnarray}
\label{cnn}
C(N,N)\,  =\, \,  {\rm det} \Bigl( a_{i-j} \Bigr), \qquad  
  1 \le i,\,\, j \le N       \\
C^*(N,N)\,  =\,\,  (-1)^{N} \cdot {\rm det} \Bigl(  a_{i-j-1} \Bigr), 
\qquad    1 \le i,\,\, j \le N       
\end{eqnarray}
where the $a_n$'s read  in terms of ${_2}F_1$ hypergeometric
 function for $\,  n \ge -1 $ 
\begin{eqnarray}
\label{lesan}
&&a_n \, =\,  - {\frac{(-1/2)_{n+1}}{(n+1)!}} \cdot 
 t^{n/2+1/2}\cdot {_2}F_1 \Bigl( 1/2, n+1/2; n+2; t \Bigr),\,\,\,
\end{eqnarray}
and for $ \,n \le -1$ :
\begin{eqnarray}
&&a_n \, = \,- {\frac{(1/2)_{-n-1}}{(-n-1)!}} \cdot  
t^{-n/2-1/2}\cdot {_2}F_1 \Bigl(- 1/2,- n-1/2;- n; t \Bigr),\,\,
\quad  \nonumber
\end{eqnarray}
where $\, (\alpha)_n$ is the usual Pochhammer symbol.

Introducing the second order linear differential operator :
\begin{eqnarray}
H_n \, = \,\,\, \, Dt^2 \, \, +{{1} \over {t}} \cdot Dt \, \,
  -\, 
{{n^2 \, t \, -(n+1)^2  } \over {4 \, (t-1) \, t^2}}
\end{eqnarray}
one can verify that $\, H_n(a_n)\, = \, 0$. One  sees
 that these second order linear
differential operators $\, H_n$
are all equivalent  (in the sense of the equivalence
of linear differential operators~\cite{Singer,PutSinger})  over 
$C(\sqrt{t})$.
Indeed, for consecutive $H_n$, we have
\begin{eqnarray}
\label{equiv}
&&H_{n} \cdot Z_n  \,\, = \,\, \,  \,\,   R_n  \cdot H_{n-1}
\qquad  \qquad  \quad \quad \hbox{with:} \\
\label{Zn}
&&\qquad  Z_n  \,\, =  \,  \,\, \sqrt {t} \cdot 
\left(  \left( t-1 \right)\cdot {\it Dt} \,\,
 + \,{\frac { \left( n-1 \right)\cdot  t\, +n}{2\, \, t}} \right)
\end{eqnarray}
We then find an intertwinning relation between $\, H_n$ and $\, H_{n-2}$.
Letting $\, \tilde{Z_2}$ denote the remainder of the rightdivision of
$Z_n\cdot Z_{n-1}$ by $\, H_{n-2}$, we find that 
$\,  H_n \cdot \tilde{Z_2}\, = \,\, \tilde{R_2} \cdot H_{n-2}$. Iteratively,
 we find an intertwinner between $\,H_n$ and $\,H_0$ that way (the same process
 is easily achieved the same way for negative values of $n$).
The degree in $\, \sqrt{t}$ of this intertwinner grows linearly.

It follows that all $\, H_n$ are equivalent (over $C(\sqrt{t})$)
to the second order differential operator $\, L_E$.
Actually, the second order differential operator
$\, L_E$ can be seen to be {\em nothing else but}
$\, H_n$ for $\, n \, = \, -1$. The equivalence (\ref{equiv}) remains valid
between  $\, H_n$ for  $\, n \, = \, \, 1$ and
$\, n \, = \, \, -1$, and, furthermore, the
equivalence between $\, L_E\,= \, H_n(n=-1)$ and
 $\, H_{n}(n=0) \, = \, \, L_{11}$ had been seen in~\cite{PainleveFuchs} ($L_{11}$ is
 the linear differential operator corresponding to $\, C(1, \, 1)$).

These equivalence of linear differential operators can be expressed
 on the entries $\, a_n$ (solutions of $\, H_n$):
\begin{eqnarray}
\label{recurr}
 a_n \, = \, \, \,
{\frac { \left(  \left( n-1 \right) t+n \right) }{2\, \, \sqrt {t}}} \cdot 
 a_{n-1}(t)
\, \, +\, \sqrt {t} \cdot \left( t-1 \right) \cdot   a_{n-1}^\prime(t).
\end{eqnarray}
Considering the fact that all $a_{n-i}$ satisfy a second order
 linear differential
 equation (namely $H_{n-i}(a_{n-i})=0$), we see that
 the above equivalences also imply that
\begin{eqnarray}
\label{recurr2}
a_n(t)\,  =\, \,\,\, t^{-n/2} \cdot \Bigl(p_n(t)\cdot  E(t)\, \,
  + q_n(t)\cdot  K(t) \Bigr) 
\end{eqnarray}
with $\, p_n,\, q_n$ polynomials in $\, t$.
Now, we have seen that the correlation functions could be seen 
as Toeplitz determinants 
in the $a_n$; so we recover the fact that the $C(N,N)$, and $C^*(N,N)$, are
 (homogeneous) polynomials in $E$ and $K$. This proof
 will be generalized in later sections.

\vskip .1cm 
{\bf Remark:} The fact that the diagonal correlations $ C(N,N) $ are
homogeneous polynomials of the first and second complete elliptic functions  $ \, K $
and $\, E $ is seen, here, as a simple consequence of the Toeplitz determinant
 representation and the contiguity relations for hypergeometric functions $\, a_n$.
Note that it can probably also be seen as obvious for some specialists of Painlev\'e,
 from the recurrence relations $\, N \mapsto \, N+1 $ given in
Jimbo and Miwa~\cite{jim-miw-80}, and from the work of
 Forrester and Witte~\cite{Witte}.

\section{The isotropic Ising model}
\label{isotropic}
In~\cite{PainleveFuchs} it was shown that the diagonal
 two-point correlation functions 
$\,C(N, \, N)$  satisfy Fuchsian linear differential equations
of order $\, N+1 $. Recalling the $\, \sigma(t)$ variables 
defined by (\ref{sigma-bas}) and  (\ref{sigma-haut}),
the compatibility between these order $\, N+1 $ Fuchsian
 linear differential equations
and (\ref{jimbo-miwa}), the sigma form of  Painlev\'e VI, 
 actually corresponds to polynomial relations~\cite{PainleveFuchs},
$\, P(\sigma', \, \sigma, \, t)\, = \, \, 0$, which, seen as functions
of $\, \sigma'$  and  $\, \sigma$ (seeing $ t \,$ 
 as a parameter) are  {\em algebraic curves of genus zero}. 

The fact that there are algebraic relations between $ \sigma(t) $ and $ \sigma'(t) $
for some classical solutions of the sixth Painlev\'e system, 
 can be seen as a consequence of the fact that classical 
solutions\footnote[4]{Classical solutions are functions obtained by finite numbers of 
differentiations, arithmetic calculations, substitution into Abelian functions, as well as
solving homogeneous linear differential equations~\cite{Umemura}.} are 
related by birational B\"acklund transformations
(in some Hamiltonian variables $ q, p $) to a seed solution which is itself
determined by a solution to a specific Riccati equation:  such algebraic
relations are implied for the $n$-th iterate of the B\"acklund transformation.

Let us recall the   $\, N\, = \, 2$ case detailed in~\cite{PainleveFuchs}.
The elimination of the variable  $S_2\, = \,\sigma''(t) $ between the
``generalized Riccati form'' of the Fuchsian ODE and
 (\ref{jimbo-miwa}), but seen  as a
polynomial relation between the three variables\footnote[5]{In 
the spirit of the ``differential algebra''~\cite{Ritt,Ritt1},
one performs as much algebraic geometry calculations as possible in 
the $n$-th derivative $ S_n=\, \sigma^{(n)}(t)$ considered 
as {\em independent variables}.
It is only at the last step that one recalls that there is
 some differential structure
 by imposing, for instance, that the variable $\, S_1 $ is 
actually the derivative with respect to $\, t$ of the  variable $\, S_0$.}
$S_0$, $\, S_1$ and $\, S_2$, yields an algebraic relation 
between $ S_0=\sigma(t)$ and   $ S_1=\, \sigma'(t)$ which 
reads the {\em rational curve}
\begin{eqnarray}
\label{nappe22}
&& \left (4\,S_0-3\right )\left (64\,{S_0}^{3}
-16\,\left (16\,t+1\right ){S_0}^{2} 
 +4\,\left (64\,{t}^{2}-16\,t-21\right ) \cdot S_0+45\right) \nonumber \\
&&\quad  -32\,t\left (4\,S_0-3\right )\left (t-1\right )
\left (8\,t-1-4\,S_0\right )\cdot  S_1 \,\nonumber  \\
&&\quad \quad  +256\,{t}^{2}\left (t-1\right )^{2}\cdot  S_1^{2}
\,\,\, =\, \,\,\, 0 
\end{eqnarray}
which is, actually, the compatibility condition
 between the Fuchsian linear differential equation
for $\,C(2, \, 2) $ and the non-linear differential
 equation (\ref{jimbo-miwa}). This can be
checked by eliminating $S_2$ between the derivative of (\ref{nappe22})
and  the Fuchsian linear differential equation
for $\,C(2, \, 2)$,
 or (\ref{jimbo-miwa}), to get again  (\ref{nappe22}).
This can also be checked directly by plugging
 a series expansion or an exact expression 
of $\, C(2,2)$ in (\ref{nappe22}).

Let us now consider the   $\, N\, = \, 3$ case, and the corresponding
compatibility condition between the Fuchsian linear differential equation
for $\,C(3, \, 3) $ and equation (\ref{jimbo-miwa}), the 
sigma form of Painlev\'e VI.
The compatibility condition also corresponds to a
 polynomial relation between $\, S_0$ and $\, S_1$,
and has been written in~\cite{PainleveFuchs}.

Seen as a relation between $\, S_0$ and $\, S_1$
(considering $\, t$  as a  parameter),
the corresponding algebraic curve  is again a {\em rational curve}.
It can thus be parametrized in term of two rational
 functions of a parameter $\, u$:
\begin{eqnarray}
\label{param}
&&S_0 \, = \, \,  \, {{N_S} \over {D_S}}, \qquad \qquad\hbox{where :} \\
&&D_S \, = \, \,  \,4\,{u}^{3}\, 
+8192\, \left( 26\,t+11 \right)  \left(t-1\right) \left( t-9 \right)
\,{t}^{2} \cdot  {u}^{2}\, \nonumber \\
&&\qquad +16777216\, \left( 19+68\,t+220\,{t}^{2} \right) 
 \left(t-1\right) ^{2} \left( t-9 \right)^{2}\,{t}^{4} \cdot u \nonumber \\
&&\,\qquad  +103079215104\, \left( 3-178\,t-140\,{t}^{2}+200\,{t}^{3} \right) 
 \, (t -1)^{3} \, (t-9)^{3} \,{t}^{6},   \nonumber   \\
&&N_S \, = \, \,  \,\left( 5+6\,t \right) \cdot {u}^{3}\, 
+2048\, \left(t-1\right)
 \, (t-9)  \left( 148\,{t}^{2}+268\,t+55 \right)\,{t}^{2} \cdot  {u}^{2}
 \nonumber \\
&&\,\qquad + 4194304\,(10\,t+19)  \, (116\,{t}^{2}+44\,t+5) 
 \, (t-1)^{2} \, (t-9)^{2} \,{t}^{4} \cdot u \nonumber \\
&&\,\qquad +8589934592\,  \cdot  (45-4560\,t-8192\,{t}^{2}+4640\,{t}^{3} 
 \nonumber \\
&& \qquad\qquad \qquad \qquad  \qquad  +2800\,{t}^{4})
 \cdot  (t-1)^{3} \, (t-9)^{3}\,{t}^{6}    ,  \nonumber  \\
&&S_1 \, = \, \, \, {{ 4 \cdot W_1 \cdot W_2} 
\over { (t-1) \cdot D_S^2 }}, \qquad \qquad \hbox{where :} \\
&&W_1 \, = \, \,  \,{u}^{2}\, +4096\,{t}^{2} \, (5+6\,t)  \, (t-1)
 \left( t-9 \right) \cdot u\,\nonumber \\
&&\qquad  +4194304\,{t}^{4}
 \left( 9-52\,t+20\,{t}^{2} \right)  \, (t-1)^{2} \, (t-9)^{2},\nonumber  \\
&&W_2 \, = \, \,  \, (3+5\,t) \cdot  {u}^{4}\, 
+8192\,\, (t-1) \left( t-9 \right)  \left( 46\,{t}^{2}+51\,t-9 \right)
 \,{t}^{2} \cdot {u}^{3}\, \nonumber \\
&&\quad - 25165824\cdot \left(165+239\,t-568\,{t}^{2}-420\,{t}^{3} \right) 
 \, (t-1)^{2} \, (t-9)^{2} \,{t}^{4}\cdot {u}^{2} \nonumber\\
&&\quad -34359738368 \cdot (1041+2881\,t+6642\,{t}^{2} \nonumber\\
&&\qquad  -4740\,{t}^{3} -3800\,{t}^{4})
  \, (t-1)^{3} \, (t-9)^{3} \,{t}^{6} \cdot u \nonumber\\
&&\qquad -17592186044416 \cdot  (3213-70749\,t-38176\,{t}^{2} \nonumber\\
&&\qquad \qquad +158280\,{t}^{3} -22800\,{t}^{4}
-34000\,{t}^{5})  \, (t-1)^{4} \, (t-9)^{4} \,{t}^{8}
\nonumber
\end{eqnarray}
Recalling that $\,  S_1$ is the derivative of $\,S_0$ with
 respect to $\, t$, one finds the 
following {\em Riccati relation} on the parameter $\, u$ :
\begin{eqnarray}
\label{Riccati}
&& {{d u} \over {dt}} \,\, = \,\, \, -{\frac {1}{16384}}\,
{\frac { u^{2}}{ (t-9) \, (t-1)^{2} \,  t^{3} }} \\ 
&& \qquad \, + \,{\frac {\left( 5\,t-13 \right)
 \left( 2\,t-9 \right) }{ 4\, (t-1)  \, (t-9) \, t  }} \cdot u 
\,\,\,\, \, \,  -256\, \, (t-9)  \, (100\,{t}^{2}-132\,t+9) \,t \nonumber 
\end{eqnarray}

We have similar results for any value of $\, N$, with, again,  Riccati
 relations on the corresponding 
rational parameter $\, u$.  
\vskip .2cm 

These results can be simply understood, and generalized, as follows. 
The diagonal correlation function $\, C(N, \, N)$
is a {\em homogeneous} polynomial~\cite{PainleveFuchs}
 of $\, E$ and $\, K$ (or $\, E$ and $E'$).
The variable $\, \sigma$ amounts to calculating the 
log-derivative of  $\, C(N, \, N)$. Recalling that the derivative of monomials 
of degree $\, N$ in  $\, E$ and $E'$, like $\, E^n \cdot (E')^{N-n}$, yield 
 monomials also of degree $\, N$ like 
$\, E^{n-1} \cdot (E')^{N+1-n}$, one easily sees that the
log-derivative of  $\, C(N, \, N)$ is the ratio of two homogeneous polynomials
of the same degree $\, N$, or, equivalently, rational
 functions of the ratio $\, \tau \, = \, E'/E$ (or $\, E/K$). Using the 
fact that $\, E$ is solution of a second order linear differential equation,
one can rewrite its second derivative with respect to $\, t$, namely $\, E''$,
into a linear combination of $\, E$ and $\, E'$ (or equivalently,
 $\, E$ and $\,K$).
One immediately deduces that  $\, \sigma'$, the first order 
derivative of $\, \sigma$
with respect to $\, t$, is also the ratio of two homogeneous polynomial
of the same degree $\, N$, or, equivalently, a {\em rational
 function} of the ratio $\, \tau \, = \, E'/E$ (or $\, E/K$). This means that 
any polynomial relation $\, P(\sigma, \, \sigma') \, = \, \, 0$ 
corresponding to the existence of a common solution $\, C(N, \, N)$
of (\ref{jimbo-miwa}) and of a $\, (N+1)$-th order Fuchsian linear
 differential equation, 
is necessarily parametrized rationally, and {\em is therefore of genus zero}.
The ``rational'' parameter of this  rational curve 
 $\, P(\sigma, \, \sigma') \, = \, \, 0$,
is, for instance, the ratio  $\, \tau $. Do note that this ratio satisfies 
a Riccati equation, in $\, t$, inherited from the second order
 differential equation satisfied by $\, E$ :
\begin{eqnarray}
{{d \, \tau} \over {dt}} \, = \, \,  \, 
A \, + \, B \cdot \tau \, + \, C \cdot \tau^2 \nonumber 
\end{eqnarray}

The emergence of a rational {\em curve} is, thus, a 
straight consequence of the {\em diagonal} 
two-point correlation functions  $\, C(N, \, N)$ being  
{\em homogeneous} polynomials 
in  $\, E$ and $\,K$. 

For the {\em off-diagonal isotropic} two-point correlation functions 
we have the following generalization : $\, \sigma$ and $\, \sigma'$
are both {\em rational expressions}\footnote[8]{Not 
birational :  $\, E$ and  $\, K$ 
are {\em not} rational expressions of $\, \sigma$ and $\, \sigma'$.} 
 of the two-variables $\, E$ and  $\, K$ 
(or equivalently $\, E$ and $\, E'$). Therefore, we are naturally
 led to consider {\em rational surfaces}~\cite{Sakai2},
 instead of rational curves.
Recalling, for instance, the polynomial expression~\cite{PainleveFuchs}
 of the off-diagonal two-point $\, C(1, \, 3)$,
the two $\, \sigma$ and  $\,\sigma'_s\, = \,  d \sigma/ds$ variables 
read respectively :
\begin{eqnarray}
\label{RatSurf} 
 {{ P_1(E, \, K)  +  P_3(E, \, K) } \over { Q_1(E, \, K)  +  Q_3(E, \, K) }},
 \, \, \, \,  {{ P_2(E, \, K)\, +  P_4(E, \, K) \, 
+  P_6(E, \, K)  } \over { \Bigl(
 Q_1(E, \, K) \, +  Q_3(E, \, K)\Bigr)^2  }} 
\end{eqnarray}
where $\, P_n$ and $\, Q_n$ denote homogeneous polynomials of degree
$\, n$ in $\, E$ and $\, K$. 

\subsection{The $ \, \mu$-extension  of the $\, C(N, \, N)$'s}
\label{ratiocurve}

Let us consider $\, C(2, \, 2)$: the three solutions of the corresponding 
Fuchsian differential operator
$\, L_{22}$  are respectively $\, C(2, \, 2)$,
 a solution with a $\log$ term for the $\, t \, \simeq \, 0$ expansions,
that we will denote  $\,{\cal S}_{1} $, and a solution with a $\, \log^2$ term,
that we will denote  $\,{\cal S}_{2}$.
Consider, now, a general linear combination of these three solutions of 
$\, L_{22}$, namely $\, C(2, \, 2) \,  \, \,
 + \, c_1 \cdot  {\cal S}_{1}\, \, + \, c_2 \cdot  {\cal S}_{2}$.  
Such a general solution of the Fuchsian ODE of order three 
is also a solution of the sigma form 
of Painlev\'e VI, (\ref{jimbo-miwa}), {\em if} (and only if
 for non-singular solutions)
it is a solution of (\ref{nappe22}). 
A straightforward calculation (using formal series in maple)
gives the following one-parameter family of solutions of (\ref{jimbo-miwa})
{\em as well as} $\, L_{22}$ :
\begin{eqnarray}
\label{C22c1c2}
&&C_{\mu}(2, \, 2) \, = \,  \,  \, \, C(2, \, 2) \,  \, \, 
+ \, c_1 \cdot {\cal S}_{1} \, \, + \, c_2 \cdot {\cal S}_{2}, \qquad \hbox{with:}
\end{eqnarray} 
\begin{eqnarray}
&&c_1 \, = \, \, \,{\frac {648 \, \, \mu}{162\, -2851 \,\mu\,
 +14255\,{\mu}^{2}}}, \qquad 
c_2 \, = \, \, \mu\,   \cdot c_1 \, \qquad \hbox{and:}\nonumber 
\end{eqnarray}
\begin{eqnarray}
\label{logS}
&& {\cal S}_{1} \, \, = \, \,  \, C(2, \, 2) \cdot \ln(t) \, + \, A_1(t), \nonumber   \\
\label{logS2}
&& {\cal S}_{2} \, \,  = \, \,  \, C(2, \, 2) \cdot \ln^2(t) \,  \\
&&\quad \quad \quad \quad  + \, \Bigl( 2 \, A_1(t)
-{{2851} \over {324}} \cdot C(2, \, 2) 
\Bigr) \cdot \ln(t) \, +  A_2(t) \nonumber
\end{eqnarray}
where  the two (holonomic) functions $\, A_1(t)$ and $\,  A_2(t)$
 have the following Laurent series expansions:
\begin{eqnarray}
 A_1(t)  =  {{2} \over {3\, t}} + {{1} \over {3}} 
+  {{151} \over { 1728}}\cdot t  +\, \cdots,   \quad \quad
 A_2(t) = {{2} \over {3\, t}}  -7 \,
-  {{1961 } \over { 1728}}\cdot t  +\, \cdots  \nonumber
\end{eqnarray}
In fact, as far as solutions of the linear 
operator $\, L_{22}$ compatible with (\ref{jimbo-miwa}) are 
concerned, since (\ref{jimbo-miwa}) bears
on  log-derivatives,  a rescaling of $C_{\mu}(2, \, 2)$ is harmless:
 we can also introduce $\, C(2, \, 2;  \, \mu) \,$
$ = \, (162\, -2851 \,\mu\, +14255\,{\mu}^{2}) \cdot C_{\mu}(2, \, 2)/162 $
which reads:
\begin{eqnarray}
 C(2, \, 2;  \, \mu) \,= \, \,\,\,  C(2, \, 2) \,  \, \, 
+ \, \mu \cdot {\cal S}_{1}^{(norm)} \, \, + \, \mu^2 \cdot {\cal S}_{2}^{(norm)}
\end{eqnarray}
where the two new  normalized solutions $\, {\cal S}_{1}^{(norm)}$
 and  $\, {\cal S}_{2}^{(norm)}$
read respectively:
\begin{eqnarray}
 4 \cdot  {\cal S}_{1} \,\, - \,  {{2851} \over {162}} \cdot C(2, \, 2), \quad \quad  \, 
 4 \cdot  {\cal S}_{2} \,\, + \,  {{14255} \over {162}} \cdot C(2, \, 2) \nonumber 
\end{eqnarray}

All these calculations are not specific of $\, N=2$ and can be generalized 
straightforwardly, for {\em any value} of
$\, N$, the only difference being that one will have to
 consider $\, N$ solutions 
$\, {\cal S}_{r}$ with their $\,\ln^r$ term~\cite{PainleveFuchs}. 
For instance, for $\, N=3$, with the formal solutions of $L_{33}$, around $\, t=0$,
written as
\begin{eqnarray}
&& {\cal S}_{0} \, \, = \, \, C(3, \, 3),   \qquad \quad 
 {\cal S}_{1} \, \, = \, \, C(3, \, 3) \cdot \ln(t) \, + \, {\cal S}_{10},  \nonumber   \\
&& {\cal S}_{2} \, \,  = \, \, C(3, \, 3) \cdot \ln^2(t) \,+\,
{\cal S}_{21}\cdot \ln(t) \, + \, {\cal S}_{20}, \\
&& {\cal S}_{3} \, \,  = \, \, C(3, \, 3) \cdot \ln^3(t) \,+\,
{\cal S}_{32}\cdot \ln^2(t) \, + \,{\cal S}_{31}\cdot \ln(t) \, +\, {\cal S}_{30} \nonumber 
\end{eqnarray}
 the linear combination
\begin{eqnarray}
 C_{\mu}(3, \, 3) \,  \,= \,  \,\,  \, \, \, C(3, \, 3)  \, \,
+ \, c_1 \cdot {\cal S}_{1} \, \, + \, c_2 \cdot {\cal S}_{2} \, \, + \, c_3 \cdot {\cal S}_{3}
\end{eqnarray} 
satisfies the nonlinear differential equation (\ref{jimbo-miwa}) with:
\begin{eqnarray}
&& c_1\, =\,\,  {\frac {810000\cdot  \left(648 -684\,\mu+11615\,{\mu}^{2} \right)\cdot \mu }
{58320000-1835320680\,\mu+22002037020\,{\mu}^{2}-99370573271\,{\mu}^{3}}} \nonumber \\
&& c_2 \,= \,\,{\frac {1944\,\mu}{648-684\,\mu
+11615\,{\mu}^{2}}} \cdot c_1, 
\quad \quad \quad \quad
c_3\, = \,\,  \mu \cdot  c_2 
\end{eqnarray}
Similarly, multiplying  $C_{\mu}(3, \, 3)$ by the denominator
 of $\, c_1$ (divided by 58320000),
and introducing well-suited normalized solutions, one can
 write a $\, \mu$-dependent solution 
of $\, L_{33}$, {\em also compatible} with (\ref{jimbo-miwa}), the sigma-form of Painlev\'e, as :
\begin{eqnarray}
 C(3, \, 3;  \, \mu) \,= \, \,\, \, C(3, \, 3) \,  \, \, 
+ \, \mu \cdot {\cal S}_{1}^{(norm)} \, \, + \, \mu^2 \cdot {\cal S}_{2}^{(norm)} 
+ \, \mu^3 \cdot {\cal S}_{3}^{(norm)} \nonumber 
\end{eqnarray}
with:
\begin{eqnarray}
&&{\cal S}_{1}^{(norm)} \, = \, \, \, 9 \cdot  {\cal S}_{1} \,\, \,
- \,  {{188819} \over {6000}} \cdot C(3, \, 3), \nonumber \\
&&{\cal S}_{2}^{(norm)} \, = \, \, \, 27 \cdot  {\cal S}_{2} \,\, \,-{{19} \over {2}} \cdot  {\cal S}_{1}
\,  + \,  {{40744513} \over {108000}} \cdot C(3, \, 3) ,  \\
&&{\cal S}_{3}^{(norm)} \, = \, \,27 \cdot  {\cal S}_{3} \,\,\, +{{11615} \over {72}} \cdot  {\cal S}_{1}
\,  - \,  {{99370573271} \over {58320000}} \cdot C(3, \, 3) \nonumber
\end{eqnarray}

This scheme will continue for {\em any value} of $\, N$.
The formal solutions of $L_{NN}$ 
which also satisfy the nonlinear differential equation (\ref{jimbo-miwa}),
can be written as:
\begin{eqnarray}
 C(N, \, N, \mu) \, = \,\, \,\,\, C(N,\, N) \,\,\,  \,
 + \, \, \sum_{j=1}^N \, \, \mu^{j} \cdot {\cal S}_{j}^{(norm)} \nonumber 
\end{eqnarray} 
where the $\, {\cal S}_{j}^{(norm)}$ are sum  of holonomic expressions 
with $\, \ln^k(t)$ terms
\begin{eqnarray}
 {\cal S}_{j}^{(norm)} \, \, = \, \,\, \,  \sum_{k=0}^j \,\,  \ln^k(t) \cdot {\cal S}_{jk}^{(norm)},
 \quad \quad \quad \quad j=\, 1,\, \cdots,\, N  \nonumber
\end{eqnarray}
where the $\, {\cal S}_{jk}^{(norm)}$ have Laurent expansions in $\, t$, around $\, t=0$.

\vskip .1cm 
Do note that such  $\, \mu$-series  with $ \ln^k t $ terms
 do not appear in the Ising correlations (i.e. $\, \mu=0$).
These  $ \, \mu$-extensions of the two-point correlation functions of the Ising model 
 are, like the $ \, \lambda$-extension of the next section, mathematical extensions of the 
Ising correlations $\, C(N, \, N)$: we do not try to give a physical content to the parameter $\, \mu$
as a  $ \, \mu$-deformation of the Ising model.

 \subsection{Towards ($\lambda$, $\, \, \mu$)-extensions of the $\, C(N, \, N)$'s}
 \label{lambda}

For a (non-linear) second-order differential equation
like (\ref{jimbo-miwa}), the sigma form of Painlev\'e VI,  corresponds to 
a {\em two-parameters} family of solutions (the ``boundary conditions''). 
In a previous paper~\cite{Holo} we underlined
 a particular one-parameter family of solutions
of (\ref{jimbo-miwa}), the so-called ``$\lambda$-extensions''
 $\, C(N, \, N; \lambda)$ that were such that their ``regular'' 
(low or high temperature) series
expansions, analytical in $\, t^{1/2}\, = \,k  = \, \, s^2$,
 were {\em actually} solutions\footnote[8]{For singled-out values 
of  $\, \lambda$, (like $\, \lambda\, = \, \, \cos(\pi m/n)$, 
$\, m, \, n$ integers), we found~\cite{Holo} that
these  $\, \lambda$-extensions are  actually solutions
 of Fuchsian linear differential equations, and
we even found that these  $\, \lambda$-extensions
 $\, C(N, \, N; \lambda)$ are {\em algebraic expressions} in $\, t$
 and, more specifically, {\em modular functions} !
} of
the sigma form of Painlev\'e VI, (\ref{jimbo-miwa}). Note that,
generically (when  $\, \lambda\, \ne \, \, \cos(\pi m/n)$),
the  $\, \lambda$-extensions
 $\, C(N, \, N; \lambda)$ {\em are not $\, D$-finite (not holonomic) anymore}.
With these $\, \lambda$-extensions we 
are performing another kind of ``deformation'' of
 $\, C(N, \, N)$: we are exploring the 
{\em analytical} (at $s=0$) {\em deformations}
 of $\, C(N, \, N)$. In contrast, with the $\, \mu$-extensions
 of the  $\, C(N, \, N)$'s of the previous subsection, we
 were exploring (in the restricted framework
of solutions of Fuchsian linear differential
 equations) ``deformations'' corresponding
 to {\em formal series} (series which are {\em not} analytic
in  $\, s$ or $\, t$, but are {\em formal series} in  $\, t$ and $\, \ln(t)$). 
 The $\, \mu$-extensions, $\, C(N, \, N;\lambda=1,  \, \mu)$,
 are analytic at $\, s=0$,  {\em only when} $\, \mu \, = \, 0$.

Of course one can ``dream'' of $\, (\lambda, \, \mu)$-extensions,
$\, C(N, \, N; \lambda, \, \mu)$,  of the diagonal
two-point correlation function  $\, C(N, \, N)$, still solutions of 
the sigma form of Painlev\'e VI, (\ref{jimbo-miwa}).
These  $\, (\lambda, \, \mu)$-extensions would be defined by {\em formal series}
that verify  (\ref{jimbo-miwa}), the sigma form of Painlev\'e VI, 
but {\em do not verify} any finite order linear differential 
equations. This more or less, amounts to considering the ``formal  series'' of
 Jimbo~\cite{Jimbo}, that we recalled in equation (5) of 
our paper~\cite{PainleveFuchs}.

\vskip .1cm

\section{The anisotropic Ising model}
\label{anis}

The previous calculations can be modified, mutatis mutandis, in the case of
the {\em anisotropic} Ising  model.
In this section, we will denote $ \, s_1 \, = \, \, \sinh(2\, K_1)$,
 $ \, s_2 \, = \, \, \sinh(2\, K_2)$,  
 $ \, c_1 \, = \, \, \cosh(2\, K_1)$ and $ \, c_2 \, = \, \, \cosh(2\, K_2)$.
We will also introduce the modulus of the elliptic
 functions parametrizing the model
 $\, k \, = \, \sinh(2\, K_1) \, \sinh(2\, K_2)$, and
 the ``anisotropy variable''
 $\, \nu \, = \, \sinh(2\, K_1)/\sinh(2\, K_2)$.
 Let us recall Montroll et al paper~\cite{mo-po-wa-63}.
The {\em off-diagonal} two-point correlation functions $\, C(N, \, M)$ are 
given, in the anisotropic case,  by  determinants
 generalizing the Toeplitz determinants
of section (\ref{first}). For the off-diagonal
 two-point correlation functions the entries $\, a_n$, 
in the corresponding  determinants, read (see (57) 
page 314 in~\cite{mo-po-wa-63}) for instance 
for the row correlation functions :
\begin{eqnarray}
\label{newan}
&&a_n \, = \, \,  \,  \,
 {{1} \over {2\, \pi}} \cdot \int^{\pi}_{-\pi} \, e^{-i\, n\, \omega}\, 
\Big({{ (z_1 \, z_2^{*} e^{i\, \omega} -1)
 \, (z_1 \, e^{i\, \omega} -z_2^{*} )} \over { 
 (e^{i\, \omega} - z_1 \, z_2^{*}) \,
 (z_2^{*} \,e^{i\, \omega} -z_1 )}}  \Bigr)^{1/2}    d\omega , \nonumber \\
&& \qquad \qquad \hbox{with :} \qquad 
\qquad  z_2^{*} \, = \,\, {{ 1 -z_2} \over { 1 +z_2  }}. 
\end{eqnarray}
where $\, z_i$ denotes the well-known high-temperature 
variables $\, \tanh(K_i)$ ($\, s_i \, 
= \, \sinh(2\, K_i)\, = \,\, 2\, z_i/(1-z_i^2)$).
These are clearly {\em holonomic functions} of  $\, z_1$ and $\, z_2$.
One can try to write the two partial differential equations satisfied by 
(\ref{newan}) in terms of $\, z_1$ and $\, z_2$. 

The simplest off-diagonal two-point correlation 
function, namely the nearest neighbour 
two-point correlation function
 $\,C(0, \, 1) \, = \,  a_0$ reads for $\, T \, > \, T_c$ 
(see eqn (4.3a) chap.8 on page 200 of~\cite{mcc-wu-73}) :
\begin{eqnarray}
\label{F00}
&&C(0, \, 1) \, = \, a_0 \, = \, \,  \\
&&\qquad \, = \, \,
2\, z_1 \, (1\, +z_2)^2 \, F_{0, \, 0} \,
 -z_1^2 \, (1\, -z_2)^2 \, F_{0, \, 1} \, 
- \, (1\, -z_2)^2 \, F_{0, \, -1}  \nonumber 
\end{eqnarray}

Since, after subsection (\ref{ellirepr}),  we have a prejudice that the 
{\em elliptic function parametrization} of the Ising
 model plays a crucial role, it is tempting to 
rewrite the previous result (\ref{F00}) (expressed in terms
 of the two variables $\, z_1$ and $\, z_2$),
in the variables  $ \, s_1$ and  $ \, s_2$ (and also 
 $ \, c_1$ and  $ \, c_2$), closer
 to the modulus of the elliptic functions of
 the Ising model. Recalling~\cite{mcc-wu-73,JournaldePhysique} and the 
modulus $\, k \, = \, \sinh(2\, K_1) \, \sinh(2\, K_2)$, 
the nearest neighbour two-point correlation function $\,C(0, \, 1)$ reads
 (see eqn (4.3a) chap.8 on page 200 of~\cite{mcc-wu-73}) 
in term of $\, \Pi_1(y, \, x)$,
the {\em Jacobi form} of the complete elliptic integral of
 the third kind~\cite{Legendre,Dieud,EDM,CRC,Abra},  and of $K(k)$, 
the complete elliptic integral of the first kind (multiplied by $2/\pi$): 
\begin{eqnarray}
\label{c01}
&&C(0, \, 1) \, = \,  \, \,  \,  \,  {{ c_1^2 \, c_2} \over {s_1}} \cdot 
  {{ 2} \over {\pi}} \,  \Pi_1(s_1^2, \, k^2) \, 
\,\, - \, {{  c_2} \over {s_1}} \cdot  \, K(k^2)  \\ 
&& \qquad \hbox{with:} \qquad \qquad  \Pi_1(y, \, x) 
\,\,=\, \, \Pi(-y, \, x) \nonumber 
\end{eqnarray}
where the  complete elliptic integral of the third kind
 $\, \Pi(y, \, x)$ reads :
\begin{eqnarray}
\label{intcomplthird}
  \Pi(y, \, x) \, = \, \, \, \int_{0}^{1} \,
 {{ du} \over {(1\,-y\, u^2)\,  \sqrt{(1\,-u^2)(1\,-x \, u^2) } }}.
\end{eqnarray}
Of course, for $\, C(1, \, 0)$, we have the same result 
as (\ref{c01}) where the
 index $\, 1$ and $\, 2$ have been permuted. Recalling the identity
\begin{eqnarray}
\label{holds}
&& {{2} \over {\pi}} \cdot \Bigl(\Pi_1(k \, \nu, \, k^2) 
 +  \Pi_1({{k} \over {\nu}}, \, k^2)\Bigr) \\
&& \qquad \qquad  \quad  = \, \, \, \,  K(k^2)\, 
+ \Bigl( (1 +k\, \nu)  (1+{{k} \over {\nu}})\Bigr)^{-1/2} \nonumber 
\end{eqnarray}
 one deduces, for instance, 
that  the following linear combination of $\,  C(0, \, 1)$
and  $\,  C(1, \, 0)$ is a function depending 
only\footnote[9]{In maple's notations,
 identity (\ref{holds}) amounts to verifying that 
$2/\pi\, EllipticPi(-k\,n,k)\,
 +2/\pi\, EllipticPi(-k/n,k)\, -2/\pi\, EllipticK(k)
 -((1+k\,n)\,(1+k/n))^{-1/2}$ equals zero.} on the modulus $\, k$ :
\begin{eqnarray}
\label{like}
&&c_1 \, c_2\, - \,s_1 \, c_2  \cdot C(0, \, 1)
 - \,s_2 \, c_1  \cdot C(1, \, 0)\,  \, \,\, = \, \,\, \,\,
  (1-k^2) \cdot K(k^2)   \\
&& = \,    c_1\, c_2 \,\, - c_1^2\, c_2^2 \cdot {{2} \over {\pi}} \cdot 
 \Bigl(\Pi_1(s_1^2, \, k^2)  + \Pi_1(s_2^2, \, k^2)  \Bigr) \, 
+ \,  K(k^2) \cdot 
(c_1^2\, +c_2^2). \nonumber 
\end{eqnarray}
In the isotropic limit, from (\ref{holds}) one easily 
gets $  2/\pi \cdot \Pi_1(k, \, k^2) =   1/(1+k)/2$$ +K(k^2)/2$.
 The Jacobi form $\, \Pi_1$ of the complete elliptic
 integral of the third kind thus reduces 
to  the complete elliptic integral of the first kind.

We see that the key role played, in the case of the isotropic
Ising model,  by the complete elliptic integral of the first, or second
kind $\, K$, or $\, E$, and the second order linear differential operator
$\, L_E$ (or $\, L_K$), is going to be played,
 in the anisotropic case,  by the {\em  complete elliptic integral of
the third kind}  $\, \Pi(y, \, x)$ and its associated partial differential
operators. Before going further in the generalizations 
of the calculations displayed in section (\ref{first}), and in the
search for the well-suited generalization of the sigma form of Painlev\'e VI
to the {\em anisotropic} Ising model, let us analyze,
 in some details, what is going to 
generalize the  second order linear differential operator
$\, L_E$ (or the Fuchs operator ${\cal L}$ in 
subsection (\ref{ellirepr}), or $\, L_K$),
 namely the partial differential operators 
corresponding to  $\, \Pi(y, \, x)$ or $\, \Pi_1(y, \, x)$, 
and, as mathematicians say, their ``${\cal D}$-module'' structure. 

\subsection{Revisiting the complete elliptic integral of the third kind }
\label{revisiting}
 
The complete elliptic integral of the third kind $\, \Pi(y, \, x)$,
is solution of two partial differential equations (see
 for instance~\cite{Wolf})
 corresponding to two partial differential operators 
that, nicely, depend,
 respectively, only on the derivative $\, D_x$ in $\, x$
and the derivative $\, D_y$ in $\, y$, separetely. We will denote these 
 two partial differential operators 
$\, {\cal L}_x $ and $\, {\cal L}_y $. 
They read respectively:
\begin{eqnarray}
\label{ellipticthird1norm}
&&{\cal L}_x  \, = \,\, \, Dx^{3}\, \,
+ {{1 } \over {2 }}\,{\frac { ( 11\,{x}^{2}-6\,x \, y\, -7\,x\, 
+2\,y )}{ (x-1)\,   (x-y)\,  x }} \cdot Dx^{2}\, \nonumber \\
&&\quad \quad \quad \quad \quad \quad \quad  +  {{ 3} \over { 4} }
 \,{\frac { (7\,x-y-2) }{ (x-1) 
 \, (x-y)\, x }} \cdot Dx  \, \,\,
+ {{ 3} \over { 8} }\,{\frac {1}{(x-1)
 \,(x-y)\,  x }}\nonumber 
\end{eqnarray}
and 
\begin{eqnarray}
\label{ellipticthird2}
&& {\cal L}_y \, = \, \,\,  D_y^{3}\, \, + {{ ( 8\,x\, y\,
 +8\,y -3\,x-13\,{y }^{2})} \over {2 \cdot  (y\, -1)  (x-y) \, y }} 
\cdot  D_y^{2}\,\, \, \nonumber \\
&&\qquad \qquad \quad \quad \quad 
+ {{ 2\, ( x-4\,y\, +1)} \over {  (y\, -1)  (x-y) \, y }} \cdot  D_y\,\,
 -{{1 } \over { (y\, -1)  (x-y) \, y }}.  \nonumber 
\end{eqnarray}

It is easy to see that $\, {\cal L}_x$ 
is always the product of an {\em order two} differential operator,
and an {\em order one} differential operator :
\begin{eqnarray}
\label{21}
&&{\cal L}_x \, = \,\,  \,
 {\cal L}^{(2)}_x \cdot {\cal L}^{(1)}_x  \qquad  \qquad \hbox{where:}  \\
&&{\cal L}^{(2)}_x \, = \,\,\,   \,  Dx^{2}\, \, 
+{\frac { ( 5\,{x}^{2}\,  +y\,  -3\,x\, (y+1) ) }
{ (x-1) \, (x-y)\,  x }} \cdot Dx\, 
\,\, + {{1 } \over {4 }} 
 \,{\frac {15\,x-3\,y-4}{ (x-1) \, (x-y)\, x  }},
\nonumber \\
&&{\cal L}^{(1)}_x \, = \,\, \,  \,Dx\,  + \, {{1} \over {2\, (x-y)}}.
\nonumber 
\end{eqnarray}
and that $\, {\cal L}_y$ is actually the product of
 the square of an order one operator with another
 order one operator :
\begin{eqnarray}
\label{111}
&&  {\cal L}_y \, = \,\,\, 
 \,\Bigl( {\cal L}_y ^{(1)} \Bigr)^{2} \cdot
 \, {\cal L}_y ^{(2)}  \qquad  \qquad \qquad \hbox{where:} \\
&& {\cal L}_y ^{(1)}  = \,    D_y 
+{\frac { 2 \, y\, (x+1) -x\, -3\,{y}^{2}}{ (y-1)
 \, (x-y) \, y}},
\, \, \, {\cal L}_y ^{(2)}  = \,   D_y
+ {{1 } \over {2 }} \,{\frac {x-{y}^{2}}{(y-1) \, (x-y) y}}.
\nonumber 
\end{eqnarray}
Generically  $\, {\cal L}_x$ or  $\, {\cal L}_y$ are
 only factored in simple product 
like (\ref{21}) and  (\ref{111}). They {\em are not 
direct sums} of linear differential operators.
They are direct sums of linear differential operators only
 for  $\,x\, = \, 0, \, 1, \, \infty $
or  $\,y\, = \, 0, \, 1, \, \infty $ (and, to some extent,
 $\,x\, = \,y$ which corresponds to the
isotropic limit of the Ising model). For instance for
  $\,y\, = \, \, 1$, the third order 
partial differential operator  $\, {\cal L}_x $ is the 
direct sum of $\, {\cal L}^{(2)}_x$ taken for $\, y \, =\, 1$,
with a second order operator $\, L_2$ (actually equivalent to $\, L_E$) :
\begin{eqnarray}
\label{norm1}
&& {\cal L}_x(y=1) \, = \, \, \,\, \,\,  \,
 {\cal L}_x ^{(1)} (y=1)\,\,  \oplus\, \, L_2 \qquad  \qquad 
\hbox{where :} \nonumber \\
&&L_2\, = \, \, \, \,  Dx^{2}\,  \, \,
+\,{\frac {2}{x-1}} \cdot Dx\, \,  \,
+  {{1 } \over {4 }}\,{\frac {1}{ (x-1)\, x }}.
\nonumber
\end{eqnarray}

Let us now analyze the differential module associated with 
 $\, {\cal L}_x$  and  $\, {\cal L}_y$
i.e the minimal system of partial linear differential equations whose 
 solutions are exactly the solutions that are common
to  ${\cal L}_x $ and ${\cal L}_y$.
Let us introduce the two polynomials
 $\, P_A \, = \, 8\, \left( x-1 \right) \, \left( x-y \right) \cdot  x$ and 
$\, P_B \, = \, 8\, \left( y-1 \right)\, \left( x-y \right)\cdot  y$,
and the two $\, 3 \times 3$ matrices 
 $\,{\cal A} \, = \,  A/P_A$ and  $\,{\cal B} \, = \,  B/P_B$, where :
\begin{eqnarray}
&&A \, = \, \,  
 \left[ \begin {array}{ccc}
 0  &\quad 8\,(x-y)\,(x-1) \,x  &0\\
\noalign{\medskip}0&0&8\,x(x-1) (x-y) \\
\noalign{\medskip}-3&\quad 6\, (y\, +2\, -7\,x )
 &\quad 4\, (7\,x\,-11\,{x}^{2} +6\,x\, y\, -2\,y) 
\end {array} \right],  \nonumber 
\end{eqnarray}
and where the $\, 3 \times 3$ matrix $\, B$ reads :
\begin{eqnarray}
&&   \left[ \begin {array}{ccc} 
4\,(x-y)^{2}&16\, (2\,x-y-1)\,(x-y)\,  x 
 &16\,(x-y)^{2} x (x-1) \\
\noalign{\medskip} 2\,(y-x)& \, \,  8\,(x-y)\,  (1\, -2\,x-y)
 &-8\, (x-y)\, ( x-1)\,x  \\
\noalign{\medskip} 3
&12\,(2\,x+y-1) &12\,(x-1)\,x\, 
 +8 \,(y-1)\,y 
\end {array}
 \right].  \nonumber 
\end{eqnarray}
Let us also introduce  $\, Z\, $ the vector of entries 
$\, z(\, x, \, y), \,\partial z / \partial x, \, 
 \partial^2 z / \partial x^2$
and the system:
\begin{eqnarray}
\label{system}
{{\partial Z } \over { \partial x}} \, = \,  \,\, 
{\cal A}  \cdot Z , \qquad \qquad 
{{\partial Z } \over { \partial y}} \, = \, \, \,
 {\cal B}  \cdot Z
\end{eqnarray}
   One easily verifies that the compatibility condition
 of this system (\ref{system}), namely
\begin{eqnarray}
\label{Schles}
 {\cal A} \cdot {\cal B} \, - \, {\cal B}  \cdot  {\cal A}  \,  \,\,\, 
+ {{\partial  {\cal A} } \over {\partial y}} \, \,
 - {{\partial {\cal B} } \over {\partial x}} \, \, = \, \, \, 0 
\end{eqnarray}
is actually satisfied. This can be seen as a 
Schlesinger system~\cite{Schlesinger}.
As a consequence, the two third-order operators  $\, {\cal L}_x$  and  $\, {\cal L}_y$ 
{\em are isomonodromic} (see this result of Malgrange 
for instance in Singer and Cassidy~\cite{Cassidy}). 

The Schlesinger system (\ref{Schles}) just means that the two 
operators  $\, {\cal L}_x$  and  $\, {\cal L}_y$ are compatible
and have a common solution, namely $\, \Pi(y, \, x)$, the
 complete elliptic of the third kind.
Instead of introducing $\, Z\, $ the vector of entries 
$\, z(y, \, x), \, \partial z/ \partial x, \, 
 \partial^2 z / \partial x^2$, one could 
have performed an
equivalent calculation with the vector of entries 
$\, z(y, \, x), \, \partial z / \partial x, \, 
 \partial z / \partial y$. This last calculation 
just corresponds to a change of basis for the two $ \, 3 \times 3$ matrices 
$\, A$ and $\, B$. 

Along this line we can recall the fact that the 
complete elliptic integral of the third kind $\Pi(y, \, x)$
verifies the following differential formula 
(see~\cite{Cayley,Wolf2} and also (3.107) and (3.112) 
in chap. 5 of~\cite{mcc-wu-73}) :
\begin{eqnarray}
\label{identonPi}
&& {{2} \over {\pi}} \cdot {{ \partial \Pi(y, \, x) }
 \over {\partial  y }} \, = \, \,\,
 {{ 1} \over {2\, (x-y) \, (y-1)}} \cdot \Bigl( E(x) \, \, \nonumber \\
&& \qquad \qquad \qquad \qquad
\,+ \,   {{x-y} \over {y }} \cdot K(x) \, 
+  {{ y^2-x } \over {y }} \cdot {{2} \over {\pi}} 
\cdot  \Pi(y, \, x)\Bigr), \nonumber \\
&& {{2} \over {\pi}} \cdot {{ \partial^2 \Pi(y, \, x) }
 \over {\partial  y^2 }} \, = \, \,\,
{{(5\, y -2) \, y + \, (1-4\, y)\, x } \over
 {4\, (x-y)\, (y-1)^2 \, y^2}} \cdot K(x) \, \nonumber \\
&&\qquad \qquad \quad -{{ (2\, y +1) \,x \,  
+ \, (2-5\, y)\, y  } \over {
 4\, (x-y)^2\, (y-1)^2 \, y}} \cdot E(x)\, \,\nonumber \\
&& \qquad \qquad \quad +{{3\, y^4\, + 2 \, (2-5\, y)\, x \, y 
+\, (4\, y-1)\, x^2 } \over { 4\, (x-y)^2\, (y-1)^2 \, y^2}} \cdot
 {{2} \over {\pi}} \cdot \Pi(y, \, x), \nonumber \\
&& {{2} \over {\pi}} \cdot {{ \partial \Pi(y, \, x) } \over {\partial  x }} \,
 = \,\,\, \,{{ 1} \over {2\, (y-x)}} \cdot \Bigl( {{ E(x)} \over {x-1}} 
\, + \, {{2} \over {\pi}} \cdot  \Pi(y, \, x)\Bigr), \\
&& {{2} \over {\pi}} \cdot {{ \partial^2 \Pi(y, \, x) }
 \over {\partial  x^2 }} \, = \,\, \,\,
{{ 4\, x^2\, -(y+2)\,\, x \, -y } \over {
 4\, (x-1)^2 \, (x-y)^2 \, x }} \cdot E(x) \nonumber \\
&&\, \,\qquad \qquad
 + {{ 1 } \over { 4\, (x-1) \,  (x-y) \, x}} \cdot K(x) \, \, 
+ \, {{3} \over {4 \, (x-y)^2 }} \cdot
 {{2} \over {\pi}} \cdot  \Pi(y, \, x). \nonumber
\end{eqnarray}
These relations show that the vector space spanned by 
$\, \Pi(y, \, x)$, $\,  \partial \Pi(y, \, x) /\partial  x$, 
$\,  \partial^2 \Pi(y, \, x) / \partial  x^2$,
the  vector space spanned by $\, \Pi_1(y, \, x)$,
 $\,  \partial \Pi(y, \, x) / \partial  x$, 
$\, \partial \Pi(y, \, x)/ \partial  y$,
and the  vector space spanned by $\, \Pi(y, \, x)$, $\, E(x)$,  $\, K(x)$,
{\em actually identify}. In particular one can write a 
 Schlesinger system (\ref{Schles}) of compatibility (isomonodromy) of  
$\, {\cal L}_x$  and  $\, {\cal L}_y$
 in the  $\, \Pi_1(y, \, x)$, $\, E(x)$,  $\, K(x)$ basis.
 
For fixed $\, y$ the  complete elliptic integral of the third kind
$\, \Pi(y, \, x)$ has two branch point
 $\, x \, = \, 1 \, $ and $\, x \, = \, \infty$. For fixed $\, x$ the
 complete elliptic integral of the third kind
$\, \Pi(y, \, x)$ has two branch points $\, y \, = \, 1 \, $ and
 $\, y \, = \, \infty$. The branch cuts location are complicated.
The  complete elliptic integral of the third kind $\, \Pi(y, \, x)$ has no
poles and essential singularities with respect to $\, y$, and similarly,
 no poles and essential singularities with
 respect to $\, x$. Less known is the fact
that the  complete elliptic integral of the third
 kind $\, \Pi(y, \, x)$ {\em can be represented 
through Appell functions\footnote[3]{Appell defined the
 functions in 1880, and Picard showed in 1881 
that they may all be expressed by integrals of the 
form : $\int_0^1 \, u^{\alpha} \cdot (1-u)^{\beta} \cdot (1\, -x\, u)^{\gamma}
\cdot (1\, -y\, u)^{\delta} \cdot du$. }, or
hypergeometric functions of two variables}~\cite{Wolf3}: 
\begin{eqnarray}
\label{hyperappel}
 {{2} \over {\pi}} \cdot \Pi(y, \, x) \,  \, = \, \, \,  
 F_{1} \Bigl(1/2; \, 1/2, \, 1;\,  1; \, x, \, y  \Bigr). 
\end{eqnarray}

Appell showed that these functions satisfy two simultaneous partial
 differential equations (see~\cite{Kampe}).   
Let us write these two partial differential operators~\cite{Koike,LeiYang} :
\begin{eqnarray}
\label{Appellst}
&&(1-y) \, y \cdot {{ \partial^2 } \over {\partial y^2}}\,
 + \, (1-y) \, x \cdot {{ \partial^2 } \over
 {\partial y\, \partial x}}\, \,\nonumber \\
&&\qquad \qquad + \, (c\, -(1+a+b')\, y) \cdot 
 {{ \partial } \over {\partial y}} \, \,
- \,  b' \, x \, {{ \partial } \over {\partial x}} \, -\,a\, b' , \nonumber 
\end{eqnarray}
\begin{eqnarray}
\label{Appell2st}
 &&(1-x) \, x \cdot {{ \partial^2 } \over {\partial x^2}}\,\,
 + \, (1-x) \, y \cdot {{ \partial^2 } 
\over {\partial x\, \partial y}}\, \,\nonumber \\
&&\qquad \qquad + \, (c\, -(1+a+b)\, y)
 \cdot  {{ \partial } \over {\partial x}} \, \,
- \, b\, y \, {{ \partial } \over {\partial y}} \, - a \, b\,
  \, \, \nonumber \\
&&\quad \quad  \, = \, \, \, \,  {\cal L} \, \, \, 
+\, (1-x) \cdot y \cdot {{ \partial^2 } \over 
{\partial x\, \partial y}}\, \, \,
- \, {{y} \over {2}} \, {{ \partial } \over {\partial y}}
\end{eqnarray}
where $\,  {\cal L} $ is the Fuchs second order linear differential operator
(\ref{Lfuchs}) of subsection (\ref{ellirepr}), where
 the variable $\, t$ has been changed into $\, x$.

In the particular case $\, c\, = 1$, $\, a\, = \, 1/2$, 
$\, b\, = \, 1/2$, $\, b' \, = \, 1$,
these equations\footnote[4]{Do note that Okamoto and Kimura have
given~\cite{Kimura} the linear partial differential 
equations for the classical seed solutions of
the two-variable Garnier system and its confluent degenerations, and the integral
representations of their solutions for general parameters. The Appell functions
are discussed there, and the general forms of
 Equations (\ref{Appell}) and (\ref{Appell2})  are
given.  } read : 
\begin{eqnarray}
\label{Appell}
&&{\cal A}_y\, = \, \,   (1-y) \, y \cdot
 {{ \partial^2 } \over {\partial y^2}}\,\,
 + \, (1-y) \, x \cdot {{ \partial^2 } \over {\partial y\, \partial x}}\, \, \\
&&\qquad \qquad \qquad 
\qquad+ \, (1\, -{{5} \over {2}}\, y) \cdot 
 {{ \partial } \over {\partial y}} \, \,\,
- \,  x \, {{ \partial } \over {\partial x}} \, -{{1} \over {2}},  \nonumber 
\end{eqnarray}
\begin{eqnarray}
\label{Appell2}
&&{\cal A}_x\, = \, \, (1-x) \, x \cdot 
{{ \partial^2 } \over {\partial x^2}}\,\,\,
 + \, (1-x) \, y \cdot {{ \partial^2 } \over {\partial x\, \partial y}}\, \, \\
&&\qquad \qquad \qquad \qquad + \, (1-2\, x) 
\cdot  {{ \partial } \over {\partial x}} \, \,\,
- \, {{y} \over {2}} \, {{ \partial } \over {\partial y}}
 \, -{{1} \over {4}}\,  \, \, \nonumber \\
&&\quad \quad  \, = \, \, \, \,\,   {\cal L} \, \, \, 
+\, (1-x) \cdot y \cdot {{ \partial^2 } \over {\partial x\, \partial y}}\, \, 
- \, {{y} \over {2}} \, {{ \partial } \over {\partial y}}. \nonumber
\end{eqnarray}

The complete elliptic integral of the third
 kind (\ref{intcomplthird}), $\, \Pi(y, \, x)$,
 is {\em a solution of that system} of partial differential operators 
(\ref{Appell}) and (\ref{Appell2}). 
The complete elliptic integral of the third kind
$\Pi(y, \, x) \, = \, \,  \, \, \,  
 \pi/2  \cdot F_{1} (1/2; \, 1/2, \, 1;\,  1; \, x, \, y  )$ 
is thus an {\em Appell function} associated 
with a system (\ref{Appell}), (\ref{Appell2}), 
closely linked to {\em del Pezzo surfaces}~\cite{Koike}
 and {\em Garnier systems}~\cite{Garnier,Koike,LeiYang}
 (along this line see also a set of 
very nice papers~\cite{Picard,Cohen,Matsu,Deligne,Shimura,Terada,Yama}).

Note that $\, x$ and $\, y$
  {\em are not on the same footing} : (\ref{Appell2}) can be 
seen as a deformation of $\, {\cal L}$  (or $\, L_K$ or $\, L_E$),
when, in contrast, (\ref{Appell}) can be seen as a deformation of 
the operator (which factorizes into two {\em order-one} operators):
\begin{eqnarray}
&&(1-y) \, y \cdot {{ \partial^2 } \over {\partial y^2}}\,\, \,
+ \, (1\, -{{5} \over {2}}\, y) \cdot  {{ \partial } \over {\partial y}} \, \, 
 \, -{{1} \over {2}} \\
&&\quad \quad \quad \quad  = \,\, \, \Bigl((1-y)\cdot  
y \cdot  {{ \partial } \over {\partial y}}\, + (1-2\,y) \Bigr) \cdot 
\Bigl( {{ \partial } \over {\partial y}}\, -{{ 1} \over { 2\, (1-y)}} \Bigr).
\nonumber 
\end{eqnarray}

These two new ``Appellian'' partial differential operators 
$\, {\cal A}_x$,  $\, {\cal A}_y$,
are slightly different from the ones we previously introduced, namely 
$\, {\cal L}_x$,  $\, {\cal L}_y$. The order-three
 partial differential operators 
$\, {\cal L}_x$,  $\, {\cal L}_y$
are more ``decoupled'' (just derivatives with
 respect to $\, x$, resp. $\, y$) than the 
 ``Appellian'' partial differential operators of order two
 $\, {\cal A}_x$,  $\, {\cal A}_y$
which present a ``mixed'' 
$  \partial^2 / \partial x\, \partial y$ derivative.  
Again, all these  partial differential operators
 have to be compatible. It is a 
straightforward, but slightly tedious, exercise to see 
that the compatibility of any choice of two
 partial differential operators among these four partial 
differential operators ($\, {\cal A}_x$, 
 $\, {\cal A}_y$, $\, {\cal L}_x$,  $\, {\cal L}_y$)
yields Schlesinger systems like (\ref{Schles}), and
 that these Schlesinger systems 
can be written in (at least) three different basis 
 ($\Pi_1(y, \, x)$, $\,  \partial \Pi_1(y, \, x)/\partial  x$, 
$\,  \partial^2 \Pi_1(y, \, x) / \partial  x^2$, or 
 $\, \Pi_1(y, \, x)$, $\,  \partial \Pi_1(y, \, x) /\partial  x$, 
$\, \partial \Pi_1(y, \, x) / \partial  y$, or  
$\, \Pi_1(y, \, x)$, $\, E(x)$,  $\, K(x)$).

\subsection{The Fuchsian PDE's of the anisotropic Ising model}
\label{generalis}
 
The calculations performed in section (\ref{first})
 can now be generalized, mutatis mutandis, replacing 
the central role played by the second order linear differential operator
$\, L_E$ (or $\, {\cal L}$, or $\, L_K$) by two 
of the partial  linear differential operators
 $\, {\cal A}_x$,  $\, {\cal A}_y$, 
$\, {\cal L}_x$,  $\, {\cal L}_y$. 
Similarly to (\ref{recurr2}), one should find that the  $\, a_n$ 
occurring in the entries of the determinants associated to the $\, C(N, \, M)$,
are linear combinations (with rational coefficients in $\, x$)
of  $\, \Pi_1(y, \, x)$, $\, E(x)$,   $\, K(x)$. 
The straight generalization of the calculations of section (\ref{first})
would correspond to write the partial linear differential operators
in $\, z_1$ and $\, z_2$ corresponding to the entries $\, a_n$,
given by the holonomic expressions (\ref{newan}), and find that these
 partial linear differential operators
are actually equivalent, and, thus equivalent (in the sense
 of equivalence of partial differential operators~\cite{Singer,PutSinger})
to the partial linear differential operators corresponding to $\, a_0$
given by (\ref{F00}) or (\ref{c01}). Relation (\ref{c01}) means that the
 partial linear differential operator, corresponding to $\, a_0$,
can be expressed as the direct sum of the partial
 linear differential operators 
in $\,s_1$ and  $\,s_2$ corresponding to 
$\, \Pi_1(s_1^2, \, \, s_1^2\, s_2^2)$ and
  $\, K(s_1^2\, s_2^2)$, that is, up to some change of 
variables (to get  partial linear differential operator 
in $\,s_1$ and  $\,s_2$), to direct sum of $\, {\cal L}_x$ and  $\, L_E$. 
These calculations are straightforward, but tedious: for instance, the 
two partial linear differential operators in  $\, z_1$ and $\, z_2$,
corresponding to $\, a_0$, are quite large: we
 have actually found, and checked, the 
Schlesinger relation like (\ref{Schles}) in  $\, z_1$
 and $\, z_2$ for these two 
partial linear differential operators. It is probably
 easier to see, directly, that the
relations generalizing (\ref{recurr2}) 
for off-diagonal two-point correlations
are actually verified. For instance the $\, a_n$ in (\ref{newan}),
corresponding to the row correlation functions (and beyond off-diagonal correlations), 
may be written:
\begin{eqnarray}
\label{prev}
&&a_n \, = \,\, 2/\pi \cdot p_n \cdot  \Pi_1(s_1^2, \,  k^2) \,\, + t_n \nonumber \\
&&\qquad \quad \quad +  q_n \cdot E(k^2) \, +  r_n \cdot K(k^2), \quad \quad \quad 
\, \,  k\,=\, \,s_1 s_2
\end{eqnarray}
where $\, p_n$,  $\, q_n$,   $\, r_n$ and $\, t_n$, are 
rational expressions in $\, s_1$ and $\, s_2$.  

The above relation (\ref{prev}), or similar relations
for the general {\em off-diagonal anisotropic} correlation functions, 
 imply that the 
{\em off-diagonal  anisotropic} correlation functions 
are {\em  non homogeneous polynomials}
 in\footnote[8]{Or, equivalently, of $\, \Pi_1(y, \, x)$,
its first and second derivatives $\, \Pi_1(y, \, x)'$
 and  $\, \Pi_1(y, \, x)''$,
with respect to the variable $\, x$, or, equivalently, of $\, \Pi_1(y, \, x)$
and its first derivatives with respect to $\, x$ and $\, y$.} the complete
 elliptic integral of the third kind,
 $\, \Pi_1(y, \, x)$, and of the first, and second, 
 complete elliptic integral\footnote[2]{To be rigorous in this polynomiality
demonstration, denoting  $\, \Phi_1(y, \, x)$, $\, \Phi_2(y, \, x)$ the
two other solutions of the third order
 linear differential operator $ \, {\cal L}_x$,
we can show  that one does not
have an algebraic relation between $\, \Pi_1(y, \, x)$,
  $\, \Phi_1(y, \, x)$, $\, \Phi_2(y, \, x)$
and their first order derivatives with
respect to $\, x$, hence the one-to-one identification
 between  homogeneous polynomials in
$\, \Pi_1(y, \, x)$,
  $\, \Phi_1(y, \, x)$, $\, \Phi_2(y, \, x)$
 and solutions of symmetric powers of $\, {\cal L}_x$.}    $\, K(x)$,  $\, E(x)$.

Most of the results that we found in~\cite{PainleveFuchs}
for the $\, C(N, \, N)$, or the  $\, C(N, \, M)$, in particular the fact that
 their corresponding linear differential operators 
are actually equivalent to the symmetric power of $\, L_E$, or direct sums of 
operators equivalent to the symmetric power of $\, L_E$, generalize mutatis mutandis 
to  direct sums of operators equivalent to least commun
 left multiple (LCLM) of $\, L_E$ and  
$\, {\cal A}_x$,  $\, {\cal A}_y$ or $\, {\cal L}_x$,  $\, {\cal L}_y$.
Most of the examples of such relations are, even in the 
simplest cases, too tedious and too large to be displayed
in this paper (due to length
 constraints of this special issue), so we 
 will  display them elsewhere.

\subsection{Generalization of Painlev\'e VI, Garnier system}
\label{higher}

  The sigma functions, associated with the 
diagonal two-point correlations $ \, C(N, \, N)$ 
 of the isotropic Ising model, are solutions of (\ref{jimbo-miwa}),
  the sigma form of Painlev\'e VI. This is even 
 true for their $\, \lambda$-extensions~\cite{Holo}. In
 a previous paper~\cite{Holo},
we used the  Ising model to give crystal clear
examples of the deep relations that exist between the Painlev\'e  VI equations,
 the theory of elliptic curves, the modular curves and 
an infinite number of Fuchsian
linear differential equation of order $\, N+1$. These deep relations, together
with the results, displayed in the previous sections, 
give a strong motivation to find the structures
that  generalize the Painlev\'e  VI equations in the
 case of the {\em off-diagonal} 
 two-point correlations $ \, C(N, \, M)$ for the 
 isotropic Ising model and, beyond, for
the {\em anisotropic} Ising model. 

For Painlev\'e specialists, the results we display in~\cite{PainleveFuchs}
for the isotropic Ising model can, at first sight, probably 
be seen as very special cases of affine Weyl group symmetries
 and Riccati type solutions of Painlev\'e equations. On the ``Painlev\'e side'' 
it is natural to try to generalize a one parameter sigma-form (\ref{jimbo-miwa}) 
to the most general four parameter case, or, even, to more general Garnier,
or Schlesinger, systems. On the lattice statistical mechanics side, it is tempting to
generalize the isotropic Ising model to a more general Yang-Baxter integrable
model with an elliptic parametrization (since we saw that the occurence of elliptic
 curves was a crucial point), namely the Baxter model, which can be seen
as two copies of the  anisotropic Ising model with a four spin coupling. 
In such a move to a  broader framework the ``dictionary'' between the 
 ``Painlev\'e language'' and the ``Yang-Baxter integrable models'' remains to be
done in a clean way. For instance, is there a correspondence 
between the $\, \mu$ and $\lambda$ 
parameters of our $\, \mu$ and $\, \lambda$
 extensions (see (\ref{ratiocurve}) and  (\ref{lambda})), 
some of the four parameters of the most general Painlev\'e equation, and 
the anisotropy,  or the four spin coupling, of the Baxter model ? 
The results we have obtained in~\cite{Holo} for singled-out values of  $\lambda$,
that the  $\lambda$ extensions of the two-point correlation functions of the Ising model
actually become algebraic functions (corresponding to modular curves), seem to indicate
that the parameter $\lambda$ identifies with the cosinus of the crossing parameter
(denoted $\, \eta$ in the Baxter model). 

Though it is probably too early to see
 the full ``global picture'', the generalizations of~\cite{PainleveFuchs}, that
we addessed here, seem to be a first, and necessary, step
paving the way to a deeper understanding of the anisotropic Ising model. 

Recalling the elliptic representation of Painlev\'e VI, which amounts to seeing
 Painlev\'e VI as a deformation
of the Fuchs-Painlev\'e second-order linear
 differential operator  $\, {\cal L}$ (or
equivalently $\, L_E$), it is  clear, for
 the {\em anisotropic} Ising model, that
 we are seeking for {\em a deformation of the ``Appellian system''
 corresponding to two of the four 
partial differential operators} $\, {\cal A}_x$,  $\, {\cal A}_y$ 
or $\, {\cal L}_x$,  $\, {\cal L}_y$.

The {\em Garnier systems}~\cite{Garnier} are
 isomonodromic systems providing the 
simplest, the most canonical, and natural, generalization of Painlev\'e VI.
The Garnier system depends on an integer $\, n$,
giving Painlev\'e VI for $\, n=1$. In an
 inspired note~\cite{Notes}, where the authors look for 
natural canonical generalizations of Painlev\'e VI, Enolskii et al
indicate that the $\, n=2$ Garnier system yields 
an {\em order four non-linear ODE}, that 
can rightly be considered as 
the {\em higher order Painlev\'e VI equation}~\cite{Notes}. 
This fourth order ODE is, however, a ``rather huge'' one. 
In general, Garnier systems and Schlesinger systems yield  
{\em systems of non-linear partial
 differential equations} rather than ODE's.  

With the  Appell-Picard hypergeometric
 differential operators in one variable, we are moving from the theory
of elliptic curves to {\em hyperelliptic curves}. {\em But is it really
 hyperelliptic curves or rational surfaces that should be considered} ? 
In the case of the {\em off-diagonal} two-point correlations
functions for the {\em isotropic} Ising model  we saw 
that the complete elliptic integral of the third kind, $\, \Pi_1$,
actually reduces to the complete elliptic integral
 of the first kind,
and that the price to pay to move from diagonal to  {\em off-diagonal} 
two-point correlations is a move from 
 {\em curves to surfaces} (rational curve to
rational surfaces: see (\ref{RatSurf}) in section 
(\ref{isotropic})). At the moment, it is still not clear
what is the proper generalization of the sigma-form of Painlev\'e VI 
for the  {\em off-diagonal} two-point 
correlations of the {\em isotropic} Ising model:
should we seek for a sigma-form  of the fourth order
 ODE previously mentioned~\cite{Notes}, or should
we look, even for the isotropic model, for system
 of PDE's associated with surfaces ?

We have a probably ``cleaner'' situation with the 
 {\em off-diagonal} two-point correlations of the
 {\em anisotropic} Ising model: to generalize the
 elliptic representation of  Painlev\'e VI
 we should probably look for some ``Appellian representation''
of Garnier systems, 
 namely a ``deformation theory'' of 
the partial differential operators $\, {\cal A}_x$,  $\, {\cal A}_y$.

\vskip .1cm 
\vskip .3cm 

\vskip .1cm

\textbf{Acknowledgments:}
We thank the referees for very usefull comments.
 One of us (JMM) thanks the MASCOS (Melbourne) where this work
 has been completed, and A. J. Guttmann for many 
fruitful discussions. We acknowledge the support of a PICS/CNRS grant. 
Two of us (BM and JMM) thank P.A. Pearce for an invitation to the SMFT 2007 
where this paper has been completed. BM and JMM also
thank P. Forrester and N. Witte for several illuminating and
very fruitful discussions.

\end{document}